\documentclass[12pt]{article}
\usepackage{epsfig}

\begin{document}

\begin{titlepage}

\hfill FTUV  07-0117


\vspace{1.5cm}

\begin{center}
\ \\
{\bf\large Monopolium: the key to monopoles}
\\
\date{ }
\vskip 0.70cm

Luis N. Epele$^{a}$, Huner Fanchiotti$^{a}$, Carlos A. Garc\'{\i}a
Canal$^{a}$
\\ and Vicente Vento$^{b}$

\vskip 0.30cm

{(a) \it Laboratorio de F\'{\i}sica Te\'{o}rica, Departamento de
F\'{\i}sica, IFLP \\ Facultad de Ciencias Exactas, Universidad
Nacional de La Plata
\\C.C. 67, 1900 La Plata, Argentina.}\\({\small E-mail:
garcia@venus.fisica.unlp.edu.ar })

\vskip 0.3cm

{(b) \it Departamento de F\'{\i}sica Te\'orica and Instituto de
F\'{\i}sica
Corpuscular}\\
{\it Universidad de Valencia and Consejo Superior
de Investigaciones Cient\'{\i}ficas}\\
{\it E-46100 Burjassot (Valencia), Spain} \\ ({\small E-mail:
Vicente.Vento@uv.es})
\end{center}

\vskip 1cm \centerline{\bf Abstract}

Dirac showed that the existence of one magnetic pole in the universe
could offer an explanation for the discrete nature of the electric
charge. Magnetic poles appear naturally in most Grand Unified
Theories. Their discovery would be of greatest importance for
particle physics and cosmology. The intense experimental search
carried thus far has not met with success. Moreover, if the
monopoles are very massive their production is outside the range of
present day facilities. A way out of this impasse would be if the
monopoles bind to form monopolium, a monopole- antimonopole bound
state, which is so strongly bound, that it has a relatively small
mass. Under these circumstances it could be produced with present
day facilities and the existence of monopoles could be indirectly
proven. We study the feasibility of detecting monopolium in present
and future accelerators.

\vspace{1cm}

\noindent Pacs: 14.80.Hv, 95.30.Cq, 98.70.-f, 98.80.-k


\end{titlepage}

\section{Introduction}

The theoretical justification for the existence of classical
magnetic poles, hereafter called monopoles, is that they add
symmetry to Maxwell's equations and explain charge quantization
\cite{dirac1,jackson} . Dirac formulated his theory of monopoles
considering them basically point like particles and quantum
mechanical consistency conditions lead to the so called Dirac
Quantization Condition (DQC),

\begin{equation} e \, g = \frac{N}{2} \;, \mbox{  N = 1,2,...}\;
, \end{equation}

\noindent where $e$ is the electron charge, $g$ the monopole
magnetic charge and we will use natural units $\hbar = c =1$. In
this theory the monopole mass, $m$, is a parameter, limited only by
classical reasonings to be $m
> 2 $ GeV \cite{book}. In non-Abelian gauge theories monopoles
arise as topologically stable solutions through spontaneous breaking
via the Kibble mechanism \cite{kibble}. They are allowed by most
Grand Unified Theory (GUT) models, have finite size and come out
extremely massive $m > 10^{16}$ GeV. There are also models based on
other mechanisms with masses between those two extremes
\cite{book,giacomelli,rujula}. All these monopoles satisfy Dirac's
quantization condition as a consequence of the monopolar structure
and the semiclassical quantization \cite{review}.

All the attempts to discover or produce monopoles have met with
failure. \cite{book,giacomelli,review,experiment,teva,mulhearn}.
This lack of experimental confirmation has led many physicist to
abandon the hope in their existence. A way out of this impasse is
the old idea of Dirac \cite{dirac1,zeldovich}, namely, monopoles are
not seen freely because they are confined by their strong magnetic
forces forming a bound state called monopolium \cite{hill,dubro}.

Several cosmological scenarios compatible with all cosmological
requirements\cite{kolb,ahlen,parker,adams} have been proposed
\cite{hill,sigl,blanco,vento} with the aim of making the existence
of monopolium compatible with the observation of ultrahigh-energy
cosmic rays (UHECRS) \cite{hayashida,bird}. From these cosmological
scenarios we have learned that the study of the monopolium
annihilation process  provides us with information regarding the
existence of monopoles, even if they are difficult to detect or to
produce in a free asymptotic state. This phenomenon is not a novel
feature of physics. Quark-gluon confinement describes the strong
limit of Quantum Chromodynamics, the theory of the hadronic
interactions, and their existence is proven by the detection of
jets, showers of conventional hadrons. There is however a main
difference between the two scenarios. In the monopolium case, the
elementary constituents may be separated asymptotically, when they
are orbiting far from each other, if the energy provided to the
system is high enough, while in the quark-gluon case this is not
possible. In practice, however, there is no big difference, since
due to the very high binding energies of monopolium, asymptotic
monopoles might only be found, for short periods of time, in the
center of galaxies, or clusters of galaxies, not in our
laboratories.

In the present work we aim at determining the existence and the
dynamics of monopoles in the laboratory. The philosophy behind the
present calculation is that the standard model allows for the
existence of monopoles which are spin 0 bosons. Therefore, given the
appropriate kinematics we should be able to produce them. However,
past experience has taught us that this is not feasible because,
most probably, their mass is very large and their production is
outside our present experimental capabilities. Our proposal here, is
that due to the large coupling between monopole and anti-monopole
the two bind to form a low mass monopolium state. This state can be
produce as an intermediate virtual state and we study its subsequent
decays. Thus, in an indirect way monopole physics can be revealed.

\section{Monopolium detection}

We proceed to discuss signatures of monopolium, the
monopole-antimonopole bound state, when produced in $e^+ e^-$
annihilation\footnote{The description in terms of quark-antiquark
annihilation is straightforward although complicated by the partonic
description of the real experimental probes which are hadrons.}. We
use to describe the interaction the low energy effective theory of
Ginzburg and Schiller \cite{ginzburg}. This theory is based on the
standard electroweak theory and in order to couple the monopoles to
the photon and weak bosons one considers that $m >> m_{Z_0}$,
$m_{Z_0}$ being the mass of the $Z_0$  boson, and that the monopole
interacts with the fundamental fields of the $SU(2) \otimes U(1) $
theory before symmetry breaking, i.e., with the isoscalar field {\em
B}, in the conventional notation of the standard model
\cite{standard}. In this way the photon, $\gamma$, and the weak
boson, $Z_0$, have the same coupling except for an additional
$\tan{\theta_W}$ , where $\theta_W$ is the Weinberg angle, for the
latter. The effective description is based on the one loop
approximation of the fundamental theory and therefore the effective
coupling is proportional to $ g_{eff} \sim \frac{\omega}{m}\, g$,
where $\omega$ is an energy scale which is below the monopole
production threshold, thus rendering the theory perturbative. The
dynamical scheme proposed by Ginzburg and Schiller leads to
effective couplings in a vector like theory between the monopole and
the photon \cite{ginzburg}, given by

\begin{equation} \label{geff}
g_{eff}^{\gamma} = C(J_m)\,g\,\frac{\omega}{m} =
C(J_m)\,\frac{\omega\, N}{2\,e\,m}
\end{equation}
with $C(J_m)\sim 1$, $\omega$ the photon energy, $m$ the monopole
mass and $N$ the monopole charge. The effective interaction between
the monopole and the $Z_0$ becomes

\begin{equation}
g_{eff}^{Z} = \tan(\theta_W)\, g_{eff}^{\gamma}
\end{equation}
where $\theta_W$ is the Weinberg angle and naturally here $\omega$
refers to the $Z_0$ energy. We have used the Dirac quantization
condition Eq.(1) to express the coupling in terms of the electron
charge.

We study the process

\begin{eqnarray}
e^+ e^- &  \rightarrow &  A \rightarrow   M + A^\prime \nonumber \\
& &  \;\;\;\;\;\;\;\;\;\; \hookrightarrow  B + C ,
\end{eqnarray}
shown in Fig. 1, where $A,A^\prime,B,C$ are $\gamma$'s or $Z_0$'s,
in all allowed combinations and $M$ represents a monopolium state.
We consider that the particle A' carries away the spin of the photon
and therefore $M$ represents the lowest scalar monopolium state. The
coupling arises as a consequence of the generalization of scalar
electrodynamics \cite{itzykson}.

\begin{figure}
\centerline{\epsfig{file=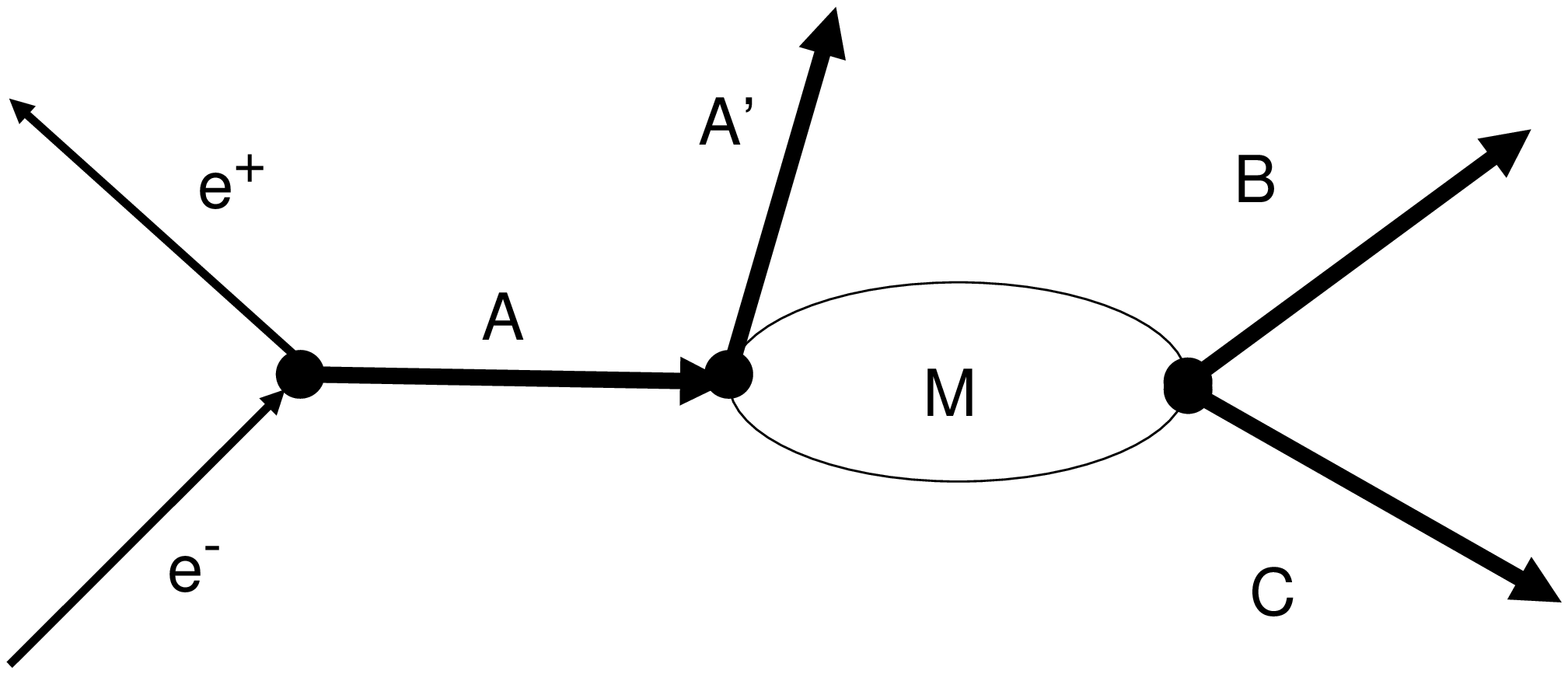,width=4cm,angle=270}}
\caption{Diagrammatic description of the reactions studied.}
\end{figure}

The standard expression for the cross section in these cases
results in

\begin{equation}
\sigma = \left(\frac{\pi}{E_e}\right)^2\,G \, \frac{4\,M ^2
\,\Gamma_{e e \gamma}\, \Gamma_{BC}}{\left(P^2 - M^2\right)^2 +
M^2\,\Gamma_M^2}
\end{equation}
Here $M$ stands for the monopolium mass, $P$ for the monopolium
four-momentum and $G$ is defined by

\[
G = \frac{2\,J_M +1}{(2\,s_e +1)\,\sqrt{(2\,s_B +1)\,(2\,s_C +1)}},
\]
where $J_M$ indicates the total angular momentum of monopolium. We
recall that the so called unitarity bound, restricts the validity of
the Ginzburg-Schiller approximation to $M < m/6$ \cite{review}. The
interesting physical situation occurs when $M\ll m$ and
consequently, from (\ref{geff}) and the fact that $\omega \sim M$,
one gets $g_{eff} \sim g (M/m) \ll 1$, which grants validity to the
perturbative approach.

We enter now the computation of the widths $\Gamma$. Taking into
account that
\[
|\emph{M}_{ee\gamma}|^2 = \left|<e^+\,e^-\,|\,\frac{1}{(q^2 -
m_A^2)}\,|\,M , A^\prime>\right|^2 = 4\,\pi\,\alpha\,g_A^2\,
g_A^{\prime\,2}\,\frac{1}{\left(q^2 -
m_A^2\right)^2}\,\left|\psi_M(0)\right|^2,
\]
where $m_A$ is the mass of the $A$ particle in Fig. 1. Here
$q=(2\,E_e,\vec{0})$. One obtains,
\begin{equation}
\Gamma_{ee\gamma} = 4
\,\alpha^2\,g_A^2\,g_{A^\prime}^2\,\frac{E_e^2}{\left(4\,E_e^2 -
m_A^2 \right)^2}\; \left|\psi_M(0)\right|^2.
\end{equation}
Note that the monopolium mass could be small and therefore we have
to keep the mass term in the denominator.

We now proceed to calculate
\[
\Gamma_{BC} = 2\,\pi\,|\emph{M}_{BC}|^2\,\rho(E_{BC})
\]
Using the standard result \cite{jauch} we obtain
\begin{equation}
\Gamma_{BC} =
\frac{8\,\pi\,g_B^2\,g_C^2}{m}\,\left|\psi_M(0)\right|^2 ,
\end{equation}
where the approximation $m\gg m_B,m_C$ has been used. Here $m_B$ and
$m_C$ represent respectively the masses of the $B$ and $C$ particles
of Fig. 1.

The cross section, generalized to include the propagation of an
unstable particle  of width $\Gamma_A$ (for us $ A = Z_0$), becomes

\begin{eqnarray} \label{c-s}
\sigma & = & \frac{2\,G}{(16)^4\,\pi^5}\,\frac{(\tan
\theta_W)^{2\zeta_0}}{\alpha^3}\,\frac{M^2}{m^{10}}\,|\psi_M(0)|^4
\nonumber \\
& & \frac{E_{A^\prime}^2 E_{B}^2\,E_C^2\,(2\,E_e)^2}{\left[(4\,E_e^2
- m_A^2)^2 + m_A^2\,\Gamma_A^2\right]\,\left[(2\,E_e-E_{A^\prime})^2
- E_{A^\prime}^2 - M^2)^2 + M^2\,\Gamma_M^2\right]}
\end{eqnarray}
where $\zeta_0$ indicates the number of $Z_0$'s present among
particles $A,\, A^\prime,\, B$ and $C$ and we have used
energy-momentum conservation at the vertex.

 In order to go ahead with the calculation, one has to
obtain the wave function corresponding to monopolium. This is done
and analyzed in the next sections.

\section{Monopolium Potential}

Our calculation of  monopolium reduces to a quantum mechanical bound
state calculation which provides us with its mass and its wave
function in the relative frame. We will use a static non
relativistic approximation and therefore our first step is to define
the potential that binds the poles to form monopolium.

We restrict our calculation to the lowest charge monopole, i.e.
$N=1$ in the Dirac condition Eq.(1). We regard the monopole as
possessing some spatial extension in line with the arguments of
Schiff and Goebel \cite{schiff,goebel}. This assumption makes the
potential energy of the monopole-antimonopole interaction
non-singular when the relative separation goes to zero.
Mathematically we describe this feature by means of an exponential
cut-off in the interaction potential,

\begin{equation} \label{v}
V(r) = - g^2\,\left(\frac{1 - \exp{(- \mu \, r)}}{r} \right).
\end{equation}

We fix the cut-off parameter $\mu$ by physical arguments.
Eq.(\ref{v}) has the following properties,

\begin{itemize}
\item [i)] $r \rightarrow \infty$
\begin{equation}
V(r) \rightarrow - \frac{g^2}{r}
\end{equation}

\item[ii)] $r \rightarrow 0$
\begin{eqnarray}
V(r) & \rightarrow & - g^2 \, \mu  + \ldots
\end{eqnarray}
\end{itemize}

When the monopole-antimonopole are closest to each other the
distance between the corresponding centers $O$ and $O^{\prime}$ is
$r_{OO^{\prime}} \sim 2\,r_m$, where $r_m$ is the pole radius. For
our estimates $r_m \sim 4\, r_{classical}$ seems reasonable since
this choice will allow the monopolium bound state to have very small
mass for strong binding. Then

\begin{equation}
\mu = 2\, \frac{m }{g^2}.
\end{equation}
Consequently, the effective potential finally becomes

\begin{equation}
V(r) = - g^2 \,\frac{1 - \exp\left(- 2
\,\frac{r}{r_{classical}}\right)}{r}.
\end{equation}
Note that with our choice, for $r \rightarrow 0$, $V(r) \rightarrow
- 2 m $. Thus, the mass of the bound state becomes the energy over
the minimum

\begin{equation}
M  = 2\,m  + E_{binding}.
\end{equation}

\begin{figure}
\centerline{\epsfig{file=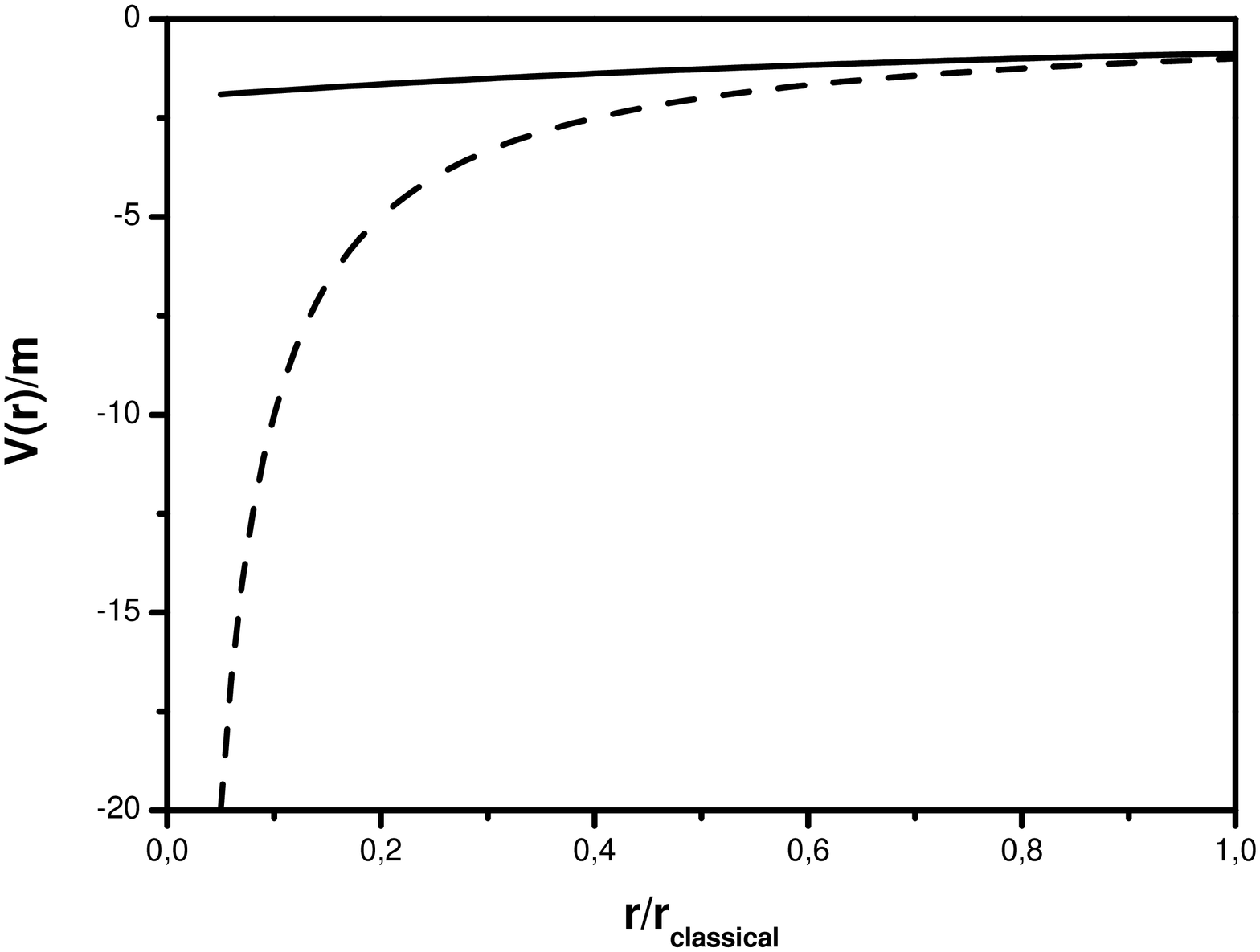,width=6.0cm,angle=270}
\hspace{0.5cm} \epsfig{file=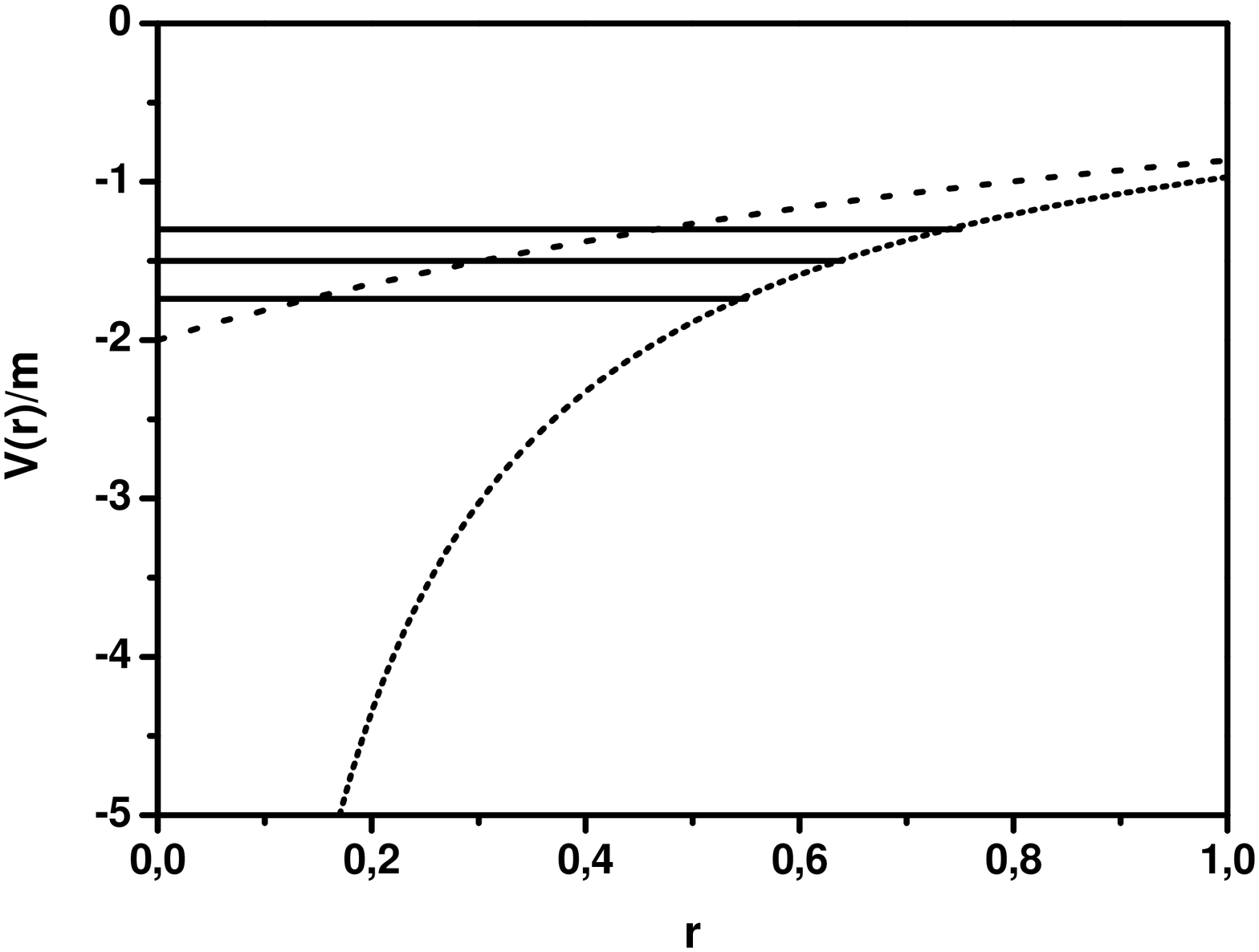,width=6.0cm,angle=270}}
\caption{The figure on the left shows the Coulomb and the Coulomb
with exponential cut-off potentials. The figure on the right shows
energy levels of the Coulomb potential used as energy levels of the
cut-off potential. Note that the lowest energy states of the cut-off
potential correspond to excited states of the Coulomb potential.}
\end{figure}

Summarizing, our analysis shows that the cut-off potential is quite
close to the Coulomb potential as long as the monopole radius, $r_m$
is greater than the classical monopole radius $ r_{classical}$.
Thus, we shall use the "magnetic" Coulomb potential (Fig. 2) as our
interaction in what follows. However, it is important to note, as
will be shown next, that the lowest energy states of the cut-off
potential correspond to excited states of the magnetic Coulomb
potential (Fig. 2.).

We use a non-relativistic approximation whose validity we will
shortly discuss. Solving the Schr\"odinger equation for monopolium
we obtain its binding energy, and therefore the mass of the system
is given by \cite{pascual}

\begin{equation}\label{M}
M = 2\,m  - \left(\frac{1}{8\,\alpha}\right)^2 \,\frac{m }{n^2}
> 0.
\end{equation}
where $\alpha = e^2 = 1/137$ and $n$ is the principal quantum
number. We see that we can reach zero mass for $n \sim 12$ and
therefore for $n > 17$ the formula is well defined and describes all
values of $M$

\[ 0 \le M\ \le 2m \].

The monopolium radius is given by

\begin{equation}\label{r}
\frac{r_M}{r_{classical}} = 48 \alpha^2 n^2.
\end{equation}
Now we introduce the size parameter $\rho = r_M/r_{classical}$.

By substituting $n^2$ from Eq.(\ref{r}) into Eq.(\ref{M}), we
obtain an equation for the monopolium mass as a function of its
size, namely

\begin{equation}
M = m  \left(2- \frac{3}{4 \rho}\right),
\end{equation}
which is plotted in Fig. 3. Although for low values of $\rho$ our
approximation becomes worse, we expect, that the soft behavior of
the wave function at the origin, allows for order of magnitude
estimates. This formula is extremely important in our development
because it transmutates principal quantum numbers of the Coulomb
potential into mass scales which are crucial to substantiate our
scenario.

\begin{figure}
\centerline{\epsfig{file=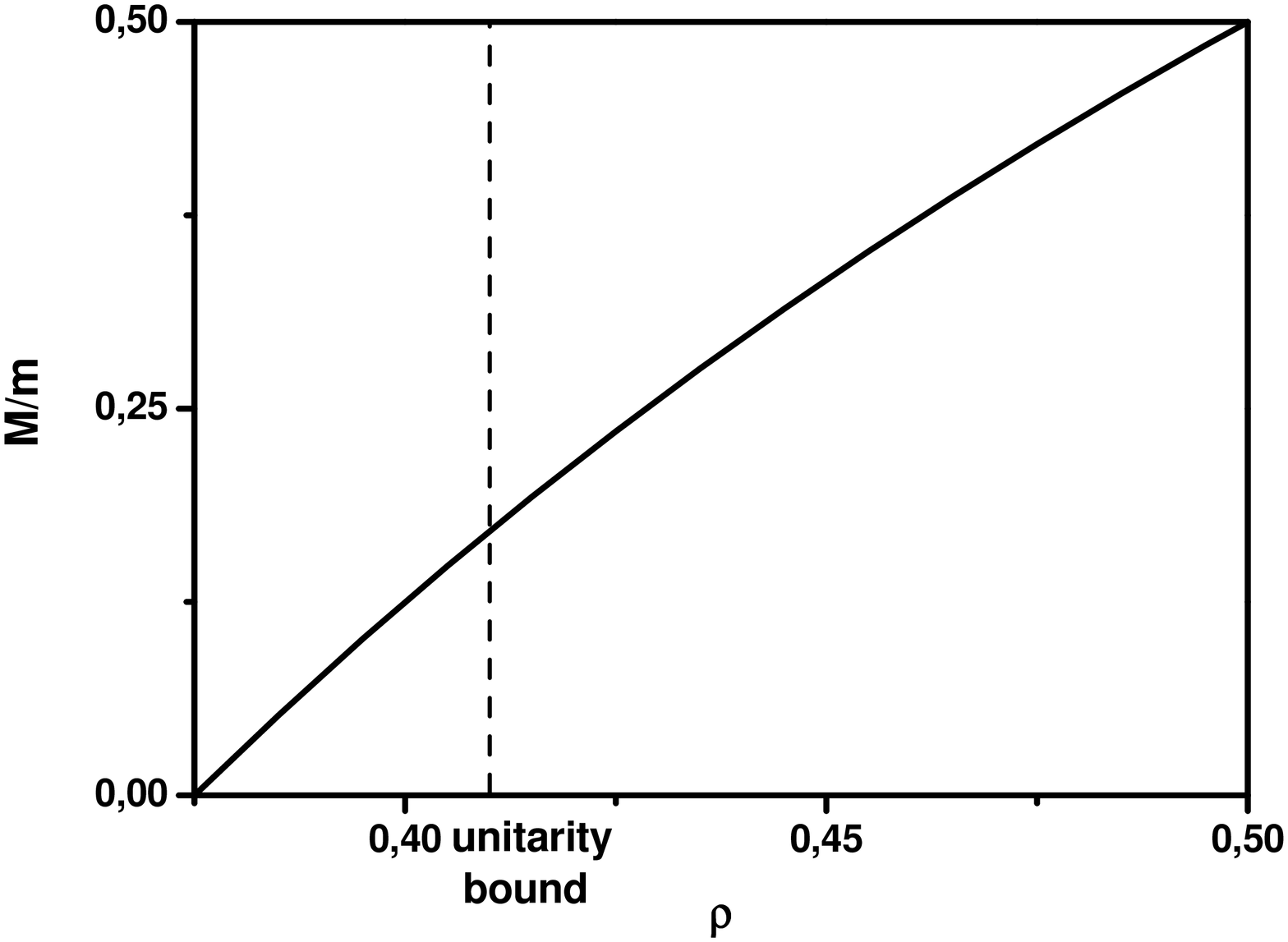,width=7cm,angle=270}}
\caption{Mass of the monopolium as a function of the size
parameter.The unitarity bound corresponding to $\rho \sim\; 0.41$ is
shown.}
\end{figure}

The perturbative expansion of Ginzburg and Schiller used in the
calculation of the production process limits, due to unitariry, the
maximum value of the ratio of $\frac{M}{m} < 1/6 $ \cite{review}. We
incorporate this external additional requirement to the bound state
calculation, which is free from it,  by showing the bound in all
relevant figures.

Before we continue, a discussion on the validity of the
non-relativistic approximation is necessary. Let us therefore
analyze how relativistic corrections will affect our calculation. As
the monopoles are spin 0 bosons one may describe the system by a
Schr\"odinger type equation and the first corrections to it do not
start at order $\beta^2$ but at order $\beta^4$ (apart from merely
the kinetic terms) \cite{yndurain}. These additional terms to order
$\beta^6$ are,

\begin{equation}
- \frac{ \Delta^4}{8 m^4} \;\; ; \; -\frac{\Delta^6}{16 m^6} \; ; \;
\frac{g^2}{32 m^5} [\Delta^2,[\Delta^2, \frac{1}{r}]].
\label{kinetic}\end{equation}
They can be treated as perturbations to the Schr\"odinger equation.
In the appendix we give detailed account of the calculation of these
corrections.

\begin{figure}
\centerline{\epsfig{file=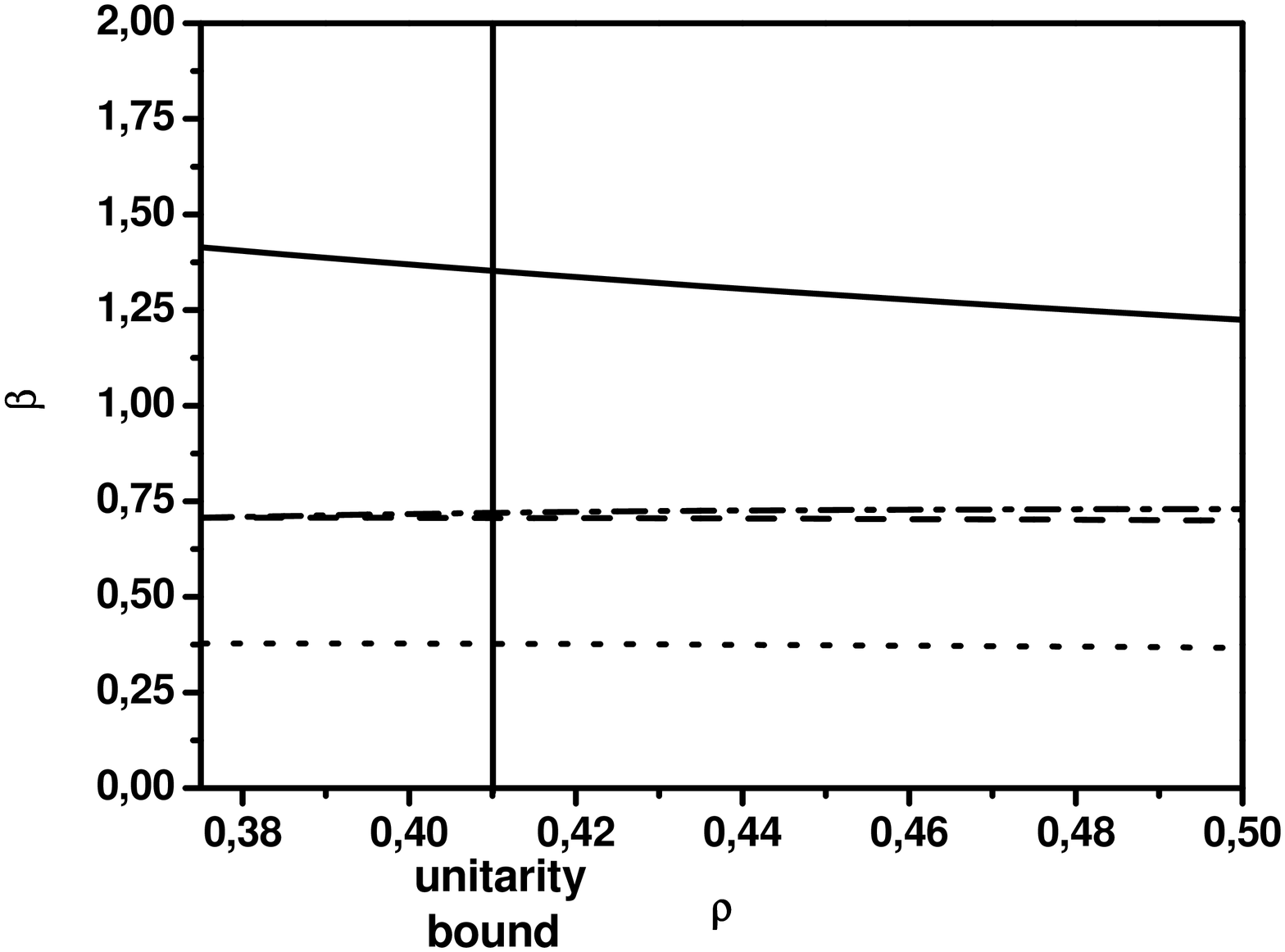,width=7cm,angle=270}}
\caption{Shown are the  non relativistic $\beta$ (solid), the
relativistic bosonic $\beta_{rel}$ (dashed), the first order
correction $\beta_{rel}$ (dot-dashed) and the ratio of an upper
bound to the order $\beta^4$ potential term to the binding energy to
order $\beta^4$ with correct kinematics (dotted). The unitarity
bound is also shown.}
\end{figure}

We show in Fig. 4 that the correction to the potential energy can be
neglected safely, for an order of magnitude estimate, if the kinetic
terms are fully taken into account. Moreover, we show that the
minimal relativistic correction, consisting in the simplest
approximation to the Klein--Gordon equation, namely in incorporating
the mass \cite{yndurain}, leads to a velocity

\begin{equation}
\beta^{(1)}_{rel}= \frac{\beta}{1 + \frac{\beta^2}{2}},
\end{equation}
which is almost indistinguishable from the more complete given by
Eq.(\ref{beta}).

Furthermore, since our potential is cut off for small values of $r$
we expect a slow down of the particles with respect to the
conventional Coulomb potential and therefore the non-relativistic
treatment is more accurate than in the pure Coulomb case. Moreover,
the calculation for the wave function at the origin is less
sensitive to the short range behavior of the potential, than the
velocity which depends on the slope of the wave function.

\section{Cross section estimates}

We proceed to use the formalism just developed to describe the
production and decay of monopolium. The analysis that follows is
physically appealing because the mass of monopolium may be chosen
small, much smaller than the monopole mass, thus detection can occur
at relatively low energies. Monopolium production is accompanied by
radiation, which is also described by the formalism. Furthermore the
calculation is easy to perform and physically understandable.

It seems therefore safe to go ahead to calculate the monopolium
decay probability as a function of $\rho$. The range of values of
$\rho$ : $3/8 < \rho < \infty$. Moreover,
\[
n = \frac{1}{4\,\alpha}\,\sqrt{\frac{\rho}{3}}
\]
and therefore, given a value of $\rho$, one can determine $n$ and
this fixes $|\psi(0)|^2$, which is what one needs for computing
the decay probability. In summary, the calculation seems to be
feasible in terms of only one mass scale, the mass of the
monopole, $m$, and one parameter, $\rho$.

Let us consider the case when monopolium is produced in its ground
state, its wave function will have $\ell=0$. Consequently
\cite{pascual}

\begin{equation}
\psi_{n,0,0} =
\frac{1}{a^{3/2}}\,N_{n,0}\,F_{n,0}\left(\frac{2\,r}{n\,a}\right)\,Y^0_0
(\Omega)
\end{equation}
with
\[
a = \frac{1}{m\,e^2} \,\,\,\,;\,\,\,\, N_{n,0} =
\frac{2}{n^2}\,\sqrt{\frac{(n-1)!}{(n!)^3}}
\]
and
\[
F_{n,0}(x) =
e^{-1/2\,x}\,L^1_{n-1}(x)\,\,\,\,;\,\,\,\,L^1_{n-1}(x) =
\sum^{n-1}_{s=0}\,(-1)^s\,\frac{(n!)^2}{(n-s-1)!\,(s+1)!\,s!}\,x^s
\]
We need $|\psi_{n,0,0}(0)|$. Then, taking into account that
\[
\lim_{x \rightarrow 0} L^1_{n-1}(x) = n\,n!\,\,\,; \,\,\, \lim_{x
\rightarrow 0} F_{n,0}(x) = n\,n!
\]
one has
\begin{equation}
|\psi_{n,0,0}(0)| =
\frac{1}{a^{3/2}}\,\frac{2}{n}\,\frac{1}{\sqrt{n}}
\end{equation}

The reduced mass of the monopolium system is $m/2$ and the Dirac
condition Eq.(1) can be written as
\[
g^2\, e^2 = \frac{1}{4}\,\,\,\,\Rightarrow \,\,\,\,
\alpha_g\,\alpha_e = \frac{1}{4}
\]
one gets
\begin{equation}
|\psi_{n,0,0}(0)| =
\frac{1}{4}\,\left(\frac{m}{2\,\alpha_e\,n}\right)^{3/2}
\end{equation}

Finally one can write the wave function in terms of the variable
$\rho$ to obtain
\begin{equation}\label{wf}
|\psi_{n,0,0}(0)| = \frac{1}{4}\,\left(\frac{2
\sqrt{3}\,m}{\sqrt{\rho}}\right)^{3/2}
\end{equation}
This is the main ingredient to be included in the expression for the
cross section that was computed before.

For spin zero monopoles and being particularly interested in $n$
large and $\ell = 0$ , one has $G\sim 1/4$.  Replacing the value of
the wave function given in (\ref{wf}), and using the relation
between $M$ and $\rho$ the cross section, Eq.(\ref{c-s}), becomes
\begin{eqnarray}
\sigma & = & \frac{1}{4^6}\frac{27}{32\,\pi^5}\,\frac{(\tan
\theta_W)^{2
N_{Z_0}}}{\alpha^3}\,\frac{M^2}{m^4}\,\left(\frac{4}{3}\,\left(2-
\frac{M}{m}\right)\right)^3\,
\nonumber \\
& & \frac{E_{A^\prime}^2\,E_{B}^2\,E_C^2\,
(2\,E_e)^2}{\left[(4\,E_e^2 - m_A^2)^2 +
(m_A^2\Gamma_A^2\right]\left[\left((2E_e- E_{A^\prime})^2 -
E_{A^\prime}^2 - M^2\right)^2 + M^2 \Gamma_M^2 \right]}.
\end{eqnarray}

We proceed to describe the case when  $A$  is a photon propagator
and the external particles $ A^\prime, B$ and $C$ are outgoing
photons. Clearly $A=Z_0$, $A^\prime = B = \gamma, C = \gamma$ also
contributes to the $3 \gamma$ cross-section. We omit it here for
simplicity, since we are just estimating the observability of the
process, not its precise magnitude.

In the case under consideration the above cross section becomes,

\begin{equation}
\sigma  = \frac{2}{4^6\pi^5\,\alpha^3}\,\frac{M^2}{m^4}\left(2-
\frac{M}{m}\right)^3\, \frac{E_{A^\prime}^2\,E_{B}^2\,E_C^2}{4 E_e^2
\; \left[\left((2 E_e- E_{A^\prime})^2 -  E_{A^\prime}^2 -
M^2\right)^2 + M^2 \Gamma_M^2 \right]}\;.
\end{equation}
It is evident from the above equation that the cross-section has a
resonant structure for

$$
E_{A^\prime} = E_e \left(1 - \left(\frac{M}{2\, E_e}\right)^2\right)
$$
In order to maximize the cross-section we will set ourselves on top
of the resonant peak and moreover will choose

$$
E_B = E_C = \frac{E_e}{2} \left(1 + \left(\frac{M}{2\,
E_e}\right)^2\right),
$$
which are the values for the energy that maximize the numerator,
$E_B^2 E_C^2$, while satisfying the conservation equations

$$
E_{A^\prime} + E_B + E_C = 2 E_e .
$$
and

$$ \vec{p}_{A^\prime} + \vec{p}_B + \vec{p}_C = \vec{0}$$
These substitutions lead finally to the formula

\begin{equation}
\sigma = \frac{2}{4^8 \pi^5 \,
\alpha^3}\left(\frac{1}{\Gamma_M}\right)^2
\left(\frac{E_e}{m}\right)^4 \left(2 - \frac{M}{m}\right)^3 \left(1
- \left(\frac{M}{2 E_e}\right)^4\right)\left(1 + \left(\frac{M}{2
E_e}\right)^2 \right).
\end{equation}
The monopolium width is dominated by the 2 $\gamma$ decay,

\begin{equation}
\Gamma_M \sim \Gamma_{BC} (E_B = E_C = M /2) = \frac{\pi}{8
\alpha^2}\left(\frac{M}{m}\right)^3\,\left(2-\frac{M}{m}\right)^{3/2}
M ,
\end{equation}
with a correction from the $Z_0$ decays which is about $30\%$ and
which for the purposes of our calculation is irrelevant. Using this
width we obtain for the cross section,

\begin{equation}
\sigma \sim \frac{2 \alpha}{4^3 \, \pi^7}
\frac{m^2}{M^4}\,\left(\left(\frac{2\,E_e}{M}\right)^4 - 1\right)
\,\left(1 + \left(\frac{M}{2 E_e}\right)^2\right).
\end{equation}

The results of our calculation are shown in Figs. 5 and 6 which we
now comment. We have two parameters in our calculation, namely the
monopole mass $m$ and the monopolium mass $M$. We use in the plots
$M$ and the ratio $M/m$. Unitarity imposes a restriction on the
latter $M/m < 0.15$ as can be seen in Fig. 3. We show data for two
values of this ratio, $0.01$, which serves for the purpose of
developing our idea and satisfies clearly the unitarity bound, and
$0.0001$, which could realize certain cosmological monopole
scenarios. The equations are sufficiently simple for any one to play
with them. In Fig. 5 we show on the left the cross section at
resonance as a function on the mass of the monopolium and for a beam
energy just above the monopolium threshold, i.e. $2\,E_e = M  +
0.001 \; GeV$. This extreme case leads to low values for the cross
section due to the vicinity of the threshold zero (see for
comparison the dotted line which is away from the threshold), but is
physically very appealing because the photon of the first vertex
carries almost no energy and we have a representation of the scalar
monopolium decay with two photons appearing back to back with an
energy of $M / 2$. In this case the photon of the first vertex
simply is present to carry away the spin of the intermediate photon
but (almost) no energy. We see that the value of the cross section
increases as $M/m$ decreases. Thus initially monopolium made of
heavier monopoles, for a fixed mass, would be easier to see, if it
were not because as shown in the right figure, its width decreases
dramatically.

\begin{figure}
\begin{center}
\epsfig{file=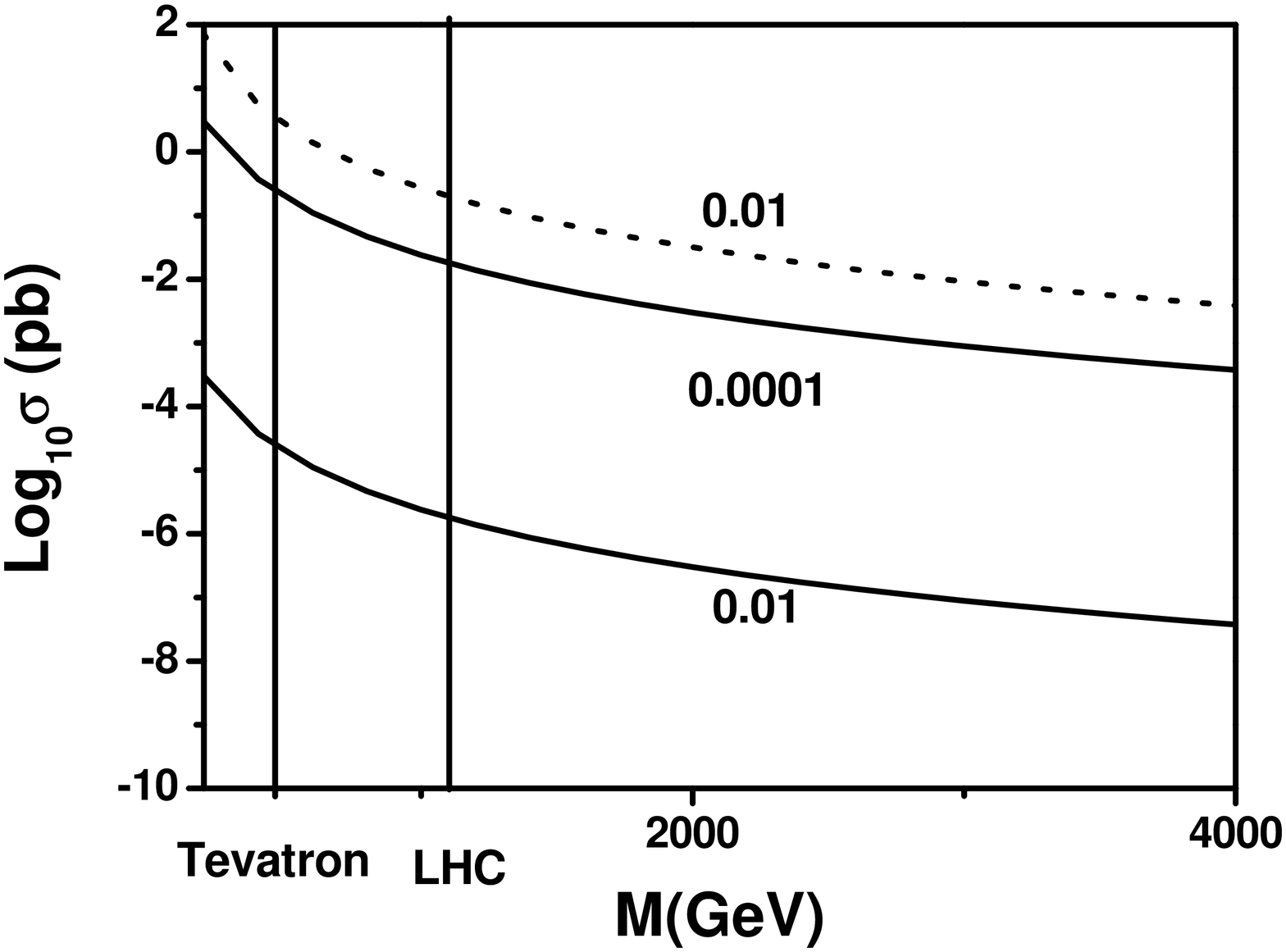,width=5.5cm,angle=270} \hspace{0.5cm}
\epsfig{file=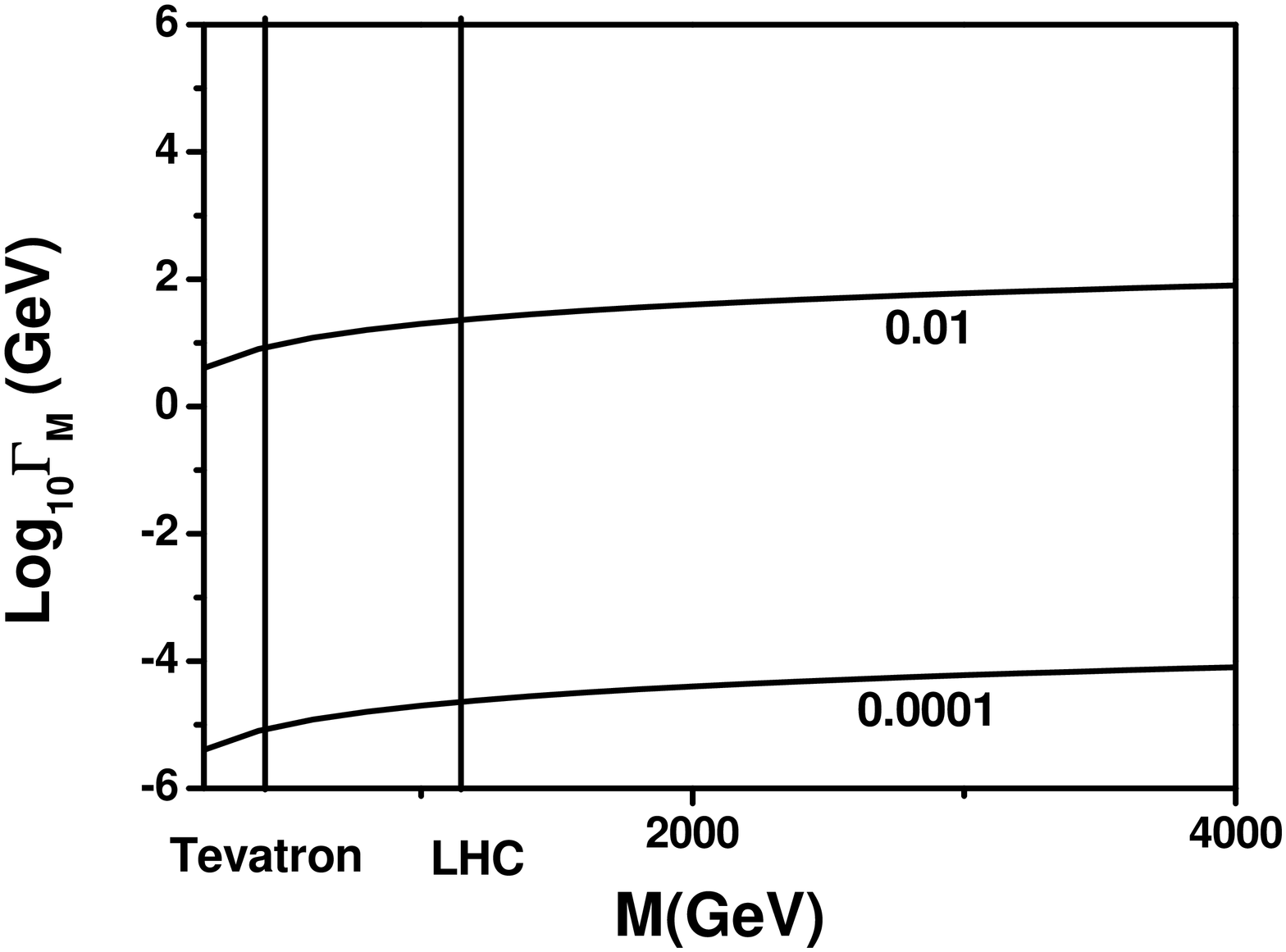,width=5.5cm,angle=270}
\end{center}
\caption{\small{ Logarithmic plot of the cross section at resonance
(in $fb$) (left) and the monopolium width (in $GeV$) (right) as a
function of the monopolium mass (in $GeV$). The vertical lines
represent the Tevatron ($M \sim 265 \, GeV$)\cite{teva} and the LHC
($M \sim 1100 \, GeV $) bounds. In the left plot the full lines have
been calculated for a beam energy slightly larger than the
monopolium mass, $2 E_e = M + 0.001$, to avoid the threshold zero
and maximize monopolium production. The dotted line corresponds to a
value of $E_e$ $100 MeV$ above the threshold. The different curves
are obtained for the shown values of $M/m = 0.01 \mbox{ or  }
0.0001$.}}
\end{figure}

We include the limits of the Tevatron and LHC, which in our way of
presenting the data are trivial. Note however, that both in the
Tevatron and in LHC, the expected processes are the inverse of the
studied here, namely the 2 $\gamma$ (and other mentioned
alternatives) of producing monopolium \cite{ginzburg,kurochin}.

Let us emphasize at this point looking at the data the interest of
what we have named a two scale scenario. The two scales are the
monopole mass and the monopolium mass. Our scenario becomes very
interesting if the two scales are very different. It is the
existence of an additional scale, which is represented in our scheme
by the parameter $M/m$, which renders the present investigation
exciting. If monopolium is a strongly bound system, it is its
relatively low mass which  limits the usefulness of the accelerators
for monopole physics and not that of the  monopole, conventionally
assumed to be very large. If monopoles could bind to an almost zero
mass bound state we could study monopole physics at relatively low
energies.

\begin{figure}
\begin{center}
\epsfig{file=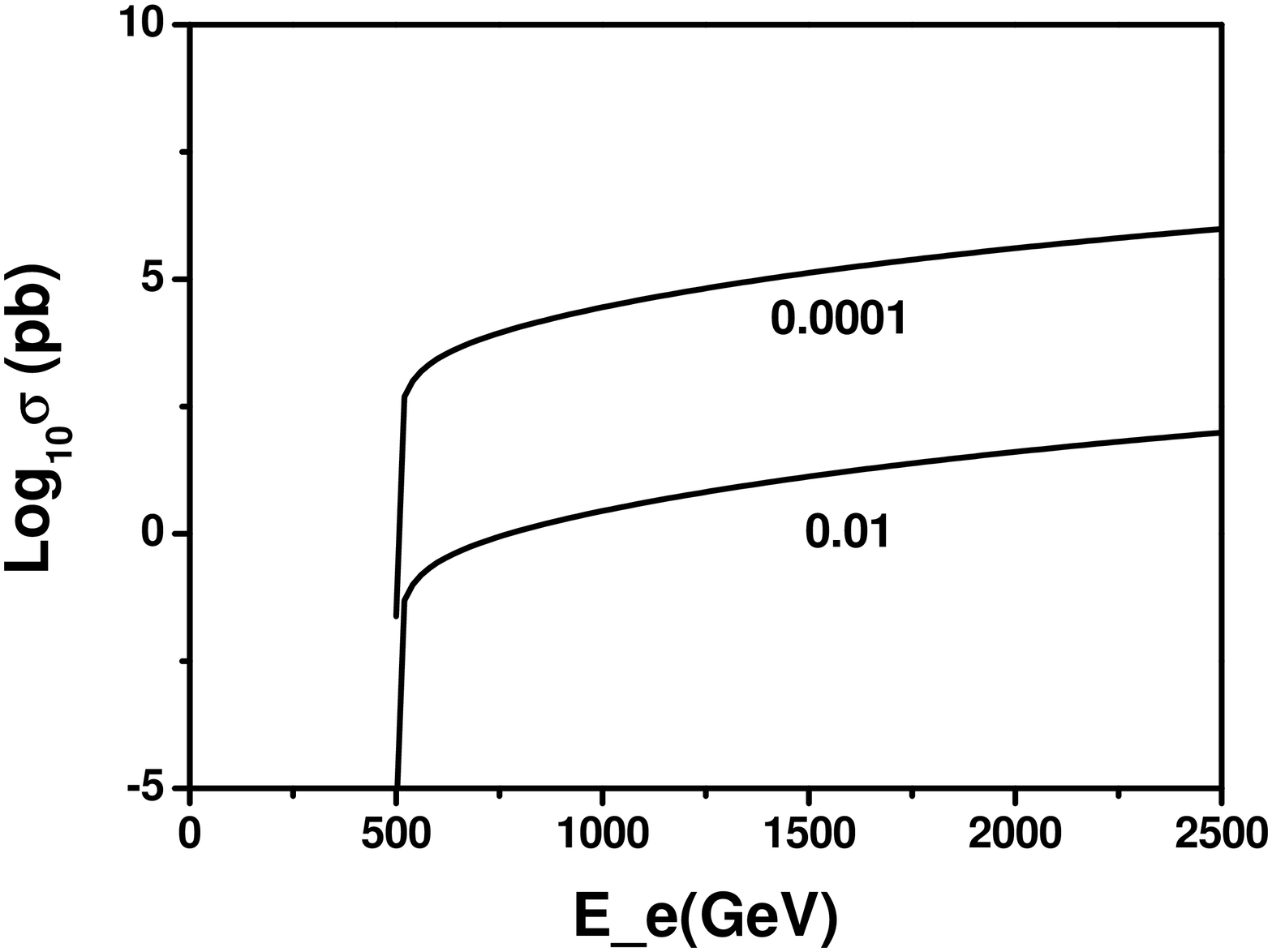,width=7.5cm,angle=270}
\end{center}
\caption{\small{ Logarithmic plot of the cross section at resonance
(in $fb$) (left) as a function of beam energy $E_e$ for a fixed mass
of the monopolium of $1000\; GeV$.  The different curves are
obtained for the shown values of $M/m = 0.01 : 0.0001$.}}
\end{figure}

In Fig. 6 we show for a fixed monopolium mass how the peak cross
section changes with the energy of the beams. We see that the
threshold effect is extremely narrow and that the cross section
jumps immediately several orders of magnitude. Thus the three photon
detection seems much more favorable than the two photon one. The
signal is clear: one photon recoiling against two others, whose
dynamics describes a resonant structure. Once the threshold effect
disappears the cross section is relatively flat with energy.

Let us summarize our findings. In our two scale scenario,

\begin{itemize}
\item [i)] the studied cross section has a resonant peak (see
Eq.(32)) at the monopolium mass $M$;

\item [ii)] the order of magnitude of the cross section is
"almost" beam energy independent, once we are away from the
threshold, and is consistent with observability in present day
machines \cite{ginzburg,kurochin,ginzburgk,brinkmann} for monopolium
masses of up to $1000 \, GeV$ (see Figs. 5 and 6) and monopole
masses only limited by the validity of our theoretical approach;

\item [iii)] a similar analysis can be carried out for $\gamma \,
Z_0$ and 2$ Z_0$ decays;

\item [iv)] a similar analysis can be carried out for hadronic
production, complicated by the inclusion of the sub-structure of
the intervening hadrons.

\item [v)] Our analysis can generalized to study the $2 \,\gamma$,
$ \gamma \, Z_0$ and $ 2 \, Z_0$ monopolium production process which
should be the dominating mechanism for LHC.

\end{itemize}

\section{Conclusions}

We have performed an investigation looking for hints of the so far
not seen monopoles. Our working assumption is that monopoles appear
strongly bound forming monopolium, a monopole-antimonopole bound
state, due to their strong electromagnetic interaction.

We develop a scenario in which monopolium is produced and
desintegrates into 2$\gamma$, $\gamma \, Z_0$ and 2$ Z_0$'s. We
detail the structure and magnitude of the first of this processes
to determine observability. We develop a two energy scale
scenario, whose

\begin{itemize}

\item[i)] low scale is governed by  monopolium and we
consider for quantitative purposes that it could be reachable by
present day machines.

\item[ii)] and whose high energy scale is governed by the monopole mass
and arises through the structure of monopolium.
$$ m >> E_e.$$

\end{itemize}

Under these circumstances we can estimate the cross section as a
function of the monopolium mass. The monopole mass is determined by
the value of the cross section and any mass is attainable, the
limitations only arising from the approximations used in our model.

Since at present we can not calculate the monopolium parameters,
$M$ and $\Gamma_M$, the experimental endeavor is not easy. There
are however some features which might simplify the task,

\begin{itemize}

\item[i)] the resonance peak of the monopolium can be found in four exit channels
3$\gamma$, $2\gamma \, Z_0$ and $ \gamma \, 2\, Z_0$'s and $ 3\,
Z_0$'s;

\item[ii)] monopolium can be produced in an excited state before
it annihilates, thus the annihilation process will be accompanied
by a Rydberg radiation spectrum;

\item[iii)] the same processes can be studied hadronically, the
only complication arising from the inclusion of the hadron
sub-structure.

\end{itemize}

The calculated values for the cross sections, corresponding to
reasonable monopolium mass scenarios, render our calculation
interesting and this line of research worth pursuing.

\section{Acknowledgement}
This work was initiated while VV was visiting the Universidad
Nacional de La Plata. He wishes to thank the members of the
department for their hospitality. VV was supported by
MCYT-FIS2004-05616-C02-01, GV-GRUPOS03/094, a travel grant of the
University of Valencia. LNE, HF and CAGC were partially supported by
CONICET and ANPCyT Argentina.

\appendix

\section{Appendix: Relativistic corrections}

Let us define the relativistic factor $\beta = v/c$ through the
equation

\begin{equation}
\beta^2 = < n \,l\, m_l\, m_s| \frac{p^2}{m^2}|n \,l \,m_l \, m_s
>.
\end{equation}
It can be easily calculated using the exact expectation value to
give \cite{pascual},

\begin{equation}
\beta = \sqrt{\frac{3}{4 \rho}}.
\end{equation}

This result, which coincides with the semiclassical treatment and
the use of Ehrenfest's theorems,
namely equating the centrifugal and Coulomb forces
\[
m\frac{v^2}{r} =
\frac{e^2}{r^2}\,\,\,\Rightarrow\,\,\,\frac{p^2}{2\,m}
=\frac{1}{2}\,\frac{e^2}{r}
\]
which leads to
\[
E = \frac{p^2}{2\,m} - \frac{e^2}{r} = \frac{p^2}{2\,m} -
\frac{p^2}{m} = - \frac{p^2}{2\,m}.
\]
This corresponds to equating the absolute values of the kinetic and
the binding energies,

\begin{equation}\label{sc}
Kinetic\,\,Energy = |Binding\,\,Energy|,
\end{equation}
which gives
\[
\frac{p^2}{m} = \left(\frac{1}{8\,\alpha} \right)^2\,\frac{m}{n^2},
\]
that gives rise, using Eq.(\ref{r}),  to
\[
\frac{p^2}{ m^2} = \frac{3}{4}\,\frac{1}{\rho}.
\]
Thus, the non-relativistic calculation is only truly valid for $\rho
> 3/4$.

One can easily incorporate the kinetic terms in Eq.(\ref{kinetic})
to the lowest order approximation leading to,

\begin{equation}
m \;(\frac{\beta^2}{2} -  \frac{\beta^4}{8} + \frac{\beta^6}{16} + .
. .)
\end{equation}
Performing the virial theorem calculation,

\begin{equation}
E_{total} = m  \;({1 + \frac{\beta^2}{2} - \frac{\beta^4}{8} +
\frac{\beta^6}{16}})
\end{equation}
and therefore the relativistic velocity turns out to be

\begin{equation}\label{beta}
\beta_{rel}= \frac{\beta}{1 + \frac{\beta^2}{2} -  \frac{\beta^4}{8}
+ \frac{\beta^6}{16}}.
\end{equation}
The non-relativistic velocity $\beta$ and the relativistic one
$\beta_{rel}$ are shown in Fig. 4. Finally the correction to the
potential compared to the binding energy becomes,

\begin{equation}
\frac{\frac{g^2}{32
m^4}<[\Delta^2,[\Delta^2,\frac{1}{r}]]>}{|E_{binding}|} <
\frac{g^2}{32}\; \frac{p ^3 }{m^3} \approx \frac{137}{128 }
\beta^3_{rel}.
\end{equation}
The upper bound is also shown in Fig. 4


\begin{thebibliography}{60}


\bibitem{dirac1} P.A.M. Dirac, Proc. Roy. Soc. {\bf A133} (1931)
60, Phys. Rev. {\bf 74} (1940) 817.

\bibitem{jackson} J. D. Jackson, Classical Electrodynamics,
de Gruyter, N.Y. (1982).

\bibitem{book} N. Craigie, G. Giacomelli, W. Nahern and Q.
Shafi, Theory and detection of magnetic monopoles in gauge theories,
World Scientific, Singapore{1986}.


\bibitem{kibble} T.W.B. Kibble, J. Phys. {\bf A 9} (1976) 1387,
Phys. Rep. {\bf 67} (1980) 183; A. Vilenkin, Phys. Rep. {\bf 121}
(1985) 263.

\bibitem{giacomelli} G. Giacomelli and L. Patrizii, hep-ex/0506014.

\bibitem{rujula} A. De R\'ujula, Nucl. Phys.{\bf B435} (1995) 257.

\bibitem{review} P.D.B. Martin Collins,A.D. Martin and E.J. Squires,
Particle Physics and cosmology, Wiley, N.Y. (1989); K. A. Milton,
Rep. Prog. Phys. {\bf 69} , 1637 (2006).

\bibitem{experiment} Review of Particle Physics, S. Eidelman et
al., Physics Letters {\bf B592 {2004} 1.}

\bibitem{teva} Reviews of Particles Properties, W-M Yao et al.,
Jour. of Phys. {\bf G33} (2006) 1.

\bibitem{mulhearn} M. J. Mulhearn, Ph.D. Thesis MIT 2004.


\bibitem{zeldovich} Ya.B Zeldovich and M. Yu. Khlopov, Phys. Lett.
{\bf 79} (1978) 239.


\bibitem{hill} C.T. Hill, Nucl. Phys. {\bf B224} (1983) 469.

\bibitem{dubro} V.K. Dubrovich. Gravitation and Cosmology, Supplement, 8, 122
(2002)

\bibitem{kolb} E.W. Kolb and M.S. Turner in {\sl The Early Universe},
Addison-Wesley, New York (1990).

\bibitem{ahlen} S. Ahlen et al., Phys. Rev. Lett. {\bf 72} (1994) 608

\bibitem{parker} E.N. Parker, Astrophys. J. {\bf 180} (1970) 383;
M.S. Turner, E.N. Parker and T. Bogdan, Phys. Rev. {\bf D26} (1982)
1296.

\bibitem{adams} F.C. Adams et al., Phys. Rev. Lett. {\bf 70} (1993)
2511.

\bibitem{sigl} P. Bhattacharjee and G. Sigl, Phys. Rev. {\bf D51} (1995) 4079.

\bibitem{blanco} J. J. Blanco-Pillado and K. D. Olum, Phys. Rev
{D60} (1999) 083001.

\bibitem{vento} V. Vento, Are monopoles hiding in monoplium?
astro-ph/0511764.


\bibitem{hayashida} N. Hayashida  et al., Phys. Rev. Lett. {\bf 73}, 3491
(1994).

\bibitem{bird} D.J. Bird  et al., Astrophys. J. {\bf 424}, 491
(1994).

\bibitem{ginzburg} I.F. Ginzburg and A. Schiller, Phys. Rev. {\bf D60},
075016 (1999); Phys. Rev. {\bf D60} 075016 (2000).

\bibitem{itzykson} C. Itzykson and B. Zuber, Quantum Field Theory,(McGraw-Hill, N.Y.
1985)

\bibitem{standard} S. Weinberg, The Quantum theory of fields (Cambridge University Press,
1996); T.P. Cheng and L.F. Lee, Theory of elementary particles
(Oxford University Press, 1982); M.E. Peskin and D.V. Schroeder,
An introduction to quantum field theory, by (HarperCollins, 1995).

\bibitem{jauch} J.M. Jauch and F. Rorhlich, The theory of
electrons and photons (Springer 1975).

\bibitem{schiff} L.I. Schiff, Phys. Rev. {\bf 160} (1967) 1257.

\bibitem{goebel} C.J. Goebel, Quanta, Essays in Theoretical
Physics, eds. P.G.O. Freund, C.J. Goebel and Y. Nambu (Chicago,
1970).

\bibitem{pascual} A. Galindo and P. Pascual, Quantum Mechanics
(Springer, 1991).

\bibitem{yndurain} F.J. Yndurain, Relativistic Quantum Mechanics
and Introduction to Field Theory (Springer 1996).


\bibitem{kurochin} Yu. Kurochin, I. Satsunkevich and Dz. Shoukavy,
Mod. Phys. Lett. {bf A21} (2006) 2873.

\bibitem{ginzburgk} I.F. Ginzburg, G.L. Kotkin, V.G. Serbo and
V. I. Telnov, Nucl. Inst. and Methods Phys. Res. {\bf 205} (1983)
47; I.F. Ginzburg, G.L. Kotkin, S.L. Panfil,  V.G. Serbo and V. I.
Telnov, Nucl. Inst. and Methods Phys. Res. {\bf 219} (1984) 5.



\bibitem{brinkmann} R. Brinkmann et al., Nucl. Inst. and Methods Phys. Res. {\bf A 406}
(1998) 13.

\end{thebibliography}
\end{document}